\documentclass[twocolumn,floatfix]{revtex4}
\usepackage{epsf,graphicx}
\usepackage{amsmath,amssymb,amsfonts}
\newcommand{\vecbm}[1]{\mbox{\boldmath#1}}
\newcommand{\vecb}[1]{\mbox{\bf#1}}
\newcommand{\cent}[1] {\begin{center}#1\end{center}}

\begin{document}
\title{Negative heat capacity at phase-separation in macroscopic systems. }
\author{D.H.E. Gross}
\affiliation{Hahn-Meitner Institute and Freie Universit{\"a}t
Berlin, Fachbereich Physik. Glienickerstr. 100; 14109 Berlin,
Germany} \email{ gross@hmi.de}
\homepage{http://www.hmi.de/people/gross/}

\begin{abstract}
Systems with long-range as well with short-range interactions should
necessarily have a convex entropy $S(E)$ at proper phase transitions of
first order, i.e. when a separation of phases occurs. Here the
microcanonical heat capacity $c(E)= -\frac{(\partial S/\partial
E)^2}{\partial^2S/\partial E^2}$ is negative. This should be observable
even in macroscopic systems when energy fluctuations with the surrounding
world can be sufficiently suppressed.
\end{abstract}
\maketitle

The argument is simple c.f.\cite{gross214}: At phase separation
the weight $e^{S(E)-E/T}$ of the configurations with energy E in
the definition of the canonical partition sum
\begin{equation}
Z(T)=\int_0^\infty{e^{S(E)-E/T}dE}\label{canonicweight}
\end{equation} becomes {\em bimodal}, at the transition temperature it has
two peaks, the ``liquid'' and the ``gas'' configurations which are
separated in energy by the latent heat. Consequently $S(E)$ must be convex
($\partial^2 S/\partial E^2>0$, like $y=x^2$) and the weight in
(\ref{canonicweight}) has a minimum at $E_{min}$ between the two pure
phases. Of course, the minimum can only be seen in the microcanonical
ensemble where the energy is controlled and its fluctuations forbidden.
Otherwise, the system would fluctuate between the two pure phases by an,
for macroscopic systems even macroscopic, energy $\Delta E\sim
E_{lat}\propto N$ of the order of the latent heat in clear contrast to the
usual assumption of the fluctuations in the canonical ensemble $\delta
E\propto \sqrt{N}$ . The heat capacity is
\begin{equation}
C_V(E_{min})=\partial E/\partial T=-~~\frac{(\partial S/\partial
E)^2}{\partial^2S/\partial E^2}<0.
\end{equation}
I.e. {\em the convexity of $S(E)$ and the negative microcanonical heat
capacity are the generic and necessary signals of  any
phase-separation\cite{gross174}}.

It is amusing that this fact, which is essential for the original purpose
of Thermodynamics to describe boiling water in steam engines, seems never
been really recognized in the past 150 years. However,  such macroscopic
energy fluctuations and the resulting negative specific heat are already
early discussed in high-energy physics by Carlitz \cite{carlitz72}.

This ``convex intruder'' in $S(E)$ with the depth $\Delta S_{surf}(E)$ has
a direct physical significance: Its depth is the surface entropy due to
constraints by the existence of the inter-phase boundary between the
droplets of the condensed phase and the gas phase and the corresponding
correlation. $\Delta S_{surf}(E)$ is directly related to the surface
tension per surface atom $N_{surf}$ of the droplets.
\begin{equation}
\sigma_{surf}/T_{tr}=\frac{\Delta S_{surf}(E_{min})}{N_{surf}}
\end{equation}

In my paper together with M.Madjet \cite{gross157} we have
compared the the values of $\Delta S_{surf}(E)$ calculated by
Monte-Carlo with the use of a realistic interaction with the
values of the surface tension of the corresponding macroscopic
system.
\begin{figure}[h]\cent{
\includegraphics*[bb = 99 57 400 286, angle=-0, width=8cm,
clip=true]{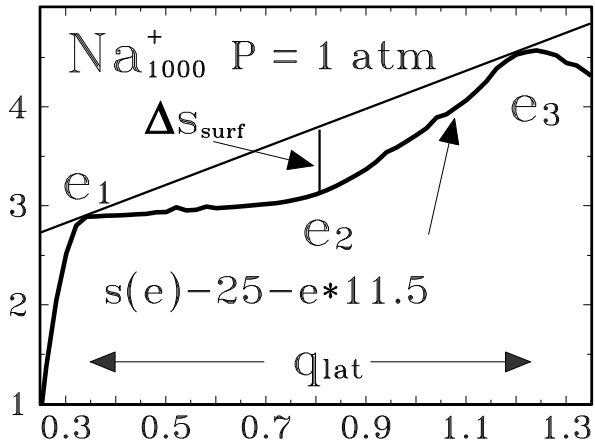}} \caption{Microcanonical Monte-Carlo
(MMMC)~\protect\cite{gross157,gross174} simulation of the entropy
  \index{entropy} $s(e)$ per atom ($e$ in eV per atom) of a system of
  $N=1000$ sodium atoms at an external pressure of 1 atm.  At the
  energy $e\leq e_1$ the system is in the pure liquid phase and at
  $e\geq e_3$ in the pure gas phase, of course with fluctuations. The
  latent heat per atom is $q_{lat}=e_3-e_1$.  \underline{Attention:}
  the curve $s(e)$ is artificially sheared by subtracting a linear
  function $25+e*11.5$ in order to make the convex intruder visible.
  {\em $s(e)$ is always a steep monotonic rising function}.  We
  clearly see the global concave (downwards bending) nature of $s(e)$
  and its convex intruder. Its depth is the entropy loss due to
  additional correlations by the interfaces. It scales $\propto
  N^{-1/3}$. From this one can calculate the surface
  tension per surface atom
  $\sigma_{surf}/T_{tr}=\Delta s_{surf}*N/N_{surf}$.  The double
  tangent (Gibbs construction) is the concave hull of $s(e)$. Its
  derivative gives the Maxwell line in the caloric curve $e(T)$ at
  $T_{tr}$. In the thermodynamic limit the intruder would disappear
  and $s(e)$ would approach the double tangent from below, not of course
  $S(E)$, which remains deeply convex: The probability of configurations with
  phase-separations is suppressed by the (infinitesimal small)
  factor $e^{-N^{2/3}}$ relative to the pure phases and the
  distribution remains {\em strictly bimodal in the canonical ensemble}.
  The region $e_1<e<e_3$ of phase separation gets lost.\label{naprl0f}}
\end{figure}
Table (\ref{table})shows the scaling behavior of $\Delta
S_{surf}(E)$ with the size $N$ of the system.
\begin{table}
\caption{Parameters of the liquid--gas transition of small
  sodium clusters (MMMC-calculation~\protect\cite{gross157,gross174}) in
  comparison with the bulk for a rising number $N$ of atoms,
  $N_{surf}$ is the average number of surface atoms (estimated here as
  $\sum{N_{cluster}^{2/3}}$) of all clusters with $N_i\geq2$ together.
  $\sigma/T_{tr}=\Delta s_{surf}*N/N_{surf}$ corresponds to the
  surface tension. Its bulk value is adjusted to agree with the
  experimental values of the $a_s$ parameter which
  we used in the liquid-drop formula for the binding energies of small
  clusters, c.f.  Brechignac et al.~\protect\cite{brechignac95}, and
  which are used in this calculation~\cite{gross174} for the individual
  clusters.\label{table}}
\begin{center}
\renewcommand{\arraystretch}{1.4}
\setlength\tabcolsep{5pt}
\begin{tabular} {|c|c|c|c|c|c|} \hline
&$N$&$200$&$1000$&$3000$&\vecb{bulk}\\ 
\hline \hline &$T_{tr} \;[K]$&$940$&$990$&$1095$&\vecb{1156}\\
\cline{2-6} &$q_{lat} \;[eV]$&$0.82$&$0.91$&$0.94$&\vecb{0.923}\\
\cline{2-6} {\bf Na}&$s_{boil}$&$10.1$&$10.7$&$9.9$&\vecb{9.267}\\
\cline{2-6} &$\Delta s_{surf}$&$0.55$&$0.56$&$0.44$&\\ \cline{2-6}
&$N_{surf}$&$39.94$&$98.53$&$186.6$&\vecbm{$\infty$}\\ \cline{2-6}
&$\sigma/T_{tr}$&$2.75$&$5.68$&$7.07$&\vecb{7.41}\\ \hline
\end{tabular}
\end{center}
\end{table}
\newpage
Roughly $\Delta S_{surf}\propto N^{2/3}$ and one may argue that this will
vanish compared to the ordinary leading volume term $S_{vol}(E)\propto N$.
However, this is not so as $S_{vol}(E)$ at energies inside the
phase-separation region (the convex intruder) is the \underline{concave
hull} of $S(E)$ (its slope gives the Maxwell construction of the caloric
curve $T(E)$). It is a straight line and its curvature $\partial^2
S_{vol}/\partial E^2\equiv 0$. Consequently for large $N$
\begin{eqnarray}
\partial^2S/\partial E^2&\sim&
\partial^2 S_{vol}/\partial E^2+\partial^2\Delta S_{surf}/\partial E^2
+\cdots \nonumber\\&\asymp& \partial^2\Delta S_{surf}/\partial E^2
\end{eqnarray} and the depth of the intruder $\Delta S_{surf}=\sigma/T_{tr}
*N_{surf}\sim N^{2/3}$ goes to infinity in the thermodynamic
limit. Of course, the ubiquitous phenomena of phase separation
exist only by this reason.

It determines the (negative) heat capacity
\begin{equation}
C_V(E_{min})= -~ \frac{(\partial S/\partial E)^2}{\partial^2S/\partial
E^2}<0.
\end{equation}

The physical (quite surprising) consequences are discussed in
\cite{gross214,gross213}.

Discussions with St.Ruffo are gratefully acknowledged.


\end{document}